\documentclass[fleqn,usenatbib]{mnras}

\usepackage[T1]{fontenc}
\usepackage{ae,aecompl}

\usepackage{graphicx}	
\usepackage{amsmath}	
\usepackage{amssymb}	

\title[GMC Properties in NGC 604]{ALMA $\mathrm{^{13}CO(J=1-0)}$ Observations of NGC 604 in M33: \ Physical Properties of Molecular Clouds}

\author[S. P. Phiri et al.]{
S.P. Phiri,$^{1}$\thanks{E-mail: spphiri@uclan.ac.uk}
J.M. Kirk,$^{1}$
D. Ward-Thompson$^{1}$ 
A.E. Sansom$^{1}$ and
G.J. Bendo$^{2}$
\\\\
$^{1}$Jeremiah Horrocks Institute, School of Natural Sciences, University of Central Lancashire, Preston, Lancashire, PR1 2HE, UK\\
$^{2}$UK ALMA Regional Centre Node, Jodrell Bank Centre for Astrophysics, Department of Physics and Astronomy, The University of\\ Manchester, Oxford Road, Manchester M13 9PL, UK
}

\date{Accepted 2021 April 19. Received 2021 March 26; in original form 2020 October 8.}

\pubyear{2020}

\begin{document}
\label{firstpage}
\pagerange{\pageref{firstpage}--\pageref{lastpage}}
\maketitle

\begin{abstract}
We present Atacama Large Millimeter/submillimeter Array (ALMA) observations of $\mathrm{^{13}CO(J=1-0)}$ line and 104 GHz continuum emission from NGC 604, a giant {H\,{\normalsize \text{II}} region} (GHR) in the nearby spiral galaxy M33. Our high spatial resolution images ($\mathrm{3.2~\arcsec~\times~2.4~\arcsec}$, corresponding to $\mathrm{13~\times~10}$ pc physical scale) allow us to detect fifteen molecular clouds. We find spatial offsets between the $\mathrm{^{13}CO}$ and 104 GHz continuum emission and also detect continuum emission near the centre of the GHR. The identified molecular clouds have  sizes ranging from 5-21 pc, linewidths of 0.3-3.0 $\mathrm{km~s^{-1}}$ and luminosity-derived masses of $\mathrm{(0.4-80.5)\times 10^3~M_{\sun}}$. These molecular clouds are in near virial equilibrium, with a spearman correlation coefficient of 0.98. The linewidth-size relationship for these clouds is offset from the corresponding relations for the Milky Way and for NGC~300, although this may be an artefact of the dendrogram process.
\end{abstract}



\section{Introduction}

Star formation occurs within cold, dense Giant Molecular Clouds (GMCs) embedded within the interstellar medium (ISM). GMCs show turbulent internal motions and are predominantly comprised of molecular hydrogen. Observations of GMCs within our own Galaxy have shown they have spatial scales of up to a hundred parsec, large scale velocity dispersions which are supersonic, and masses up to $10^6$ solar masses \citep{Heyer2015}. Three key empirical GMC scaling relations, which have become officially accepted diagnostics for the physical conditions and structure, were first identified by \citet{Larson1981}. Later studies by other authors (e.g., \citep[][and references therein]{Solomon1987,Rice2016}) have demonstrated the ubiquity of these scaling relations, commonly called Larson's relations, for Milky Way clouds. The first scaling relation is the size - linewidth relation, where the velocity line width of giant molecular clouds is proportional to the 0.5 power of the size, $\mathrm{\Delta v \varpropto R^{0.5}}$. The second relation deals with GMC's virial equilibrium{\bf ,} where gravitational potential energy and kinetic energy are in approximate equilibrium \citep{Larson1981,Solomon1987,Heyer2009,Heyer2015}. This equilibrium manifests as a direct correlation between the masses estimated from related methods, e.g. the virial mass ($M_vir$) and the luminous mass (e.g, from $\mathrm{^{13}CO}$ in our case). A final implication of the Larson scaling relationships is that the surface density of molecular is approximately constant (e.g.,$\Sigma \varpropto M/R^2 \varpropto \rho R$). This proceeds from the Third Law which showed that $\rho \propto 1/R$ where R is an estimate of its physical size of the cloud and $\rho$ is its mass volume density \citep{Larson1981}. This clear universality in cloud structure was verified in other Galactic studies \citep{Solomon1987,Heyer2009}. Some extragalactic studies have also found correlations between GMC size and mass \citep{Bolatto2008,Hughes2010}. However, \citet{Faesi2018} notes that these extragalactic observations have low sensitivities, with the majority of pixels in GMCs near the sensitivity threshold, so the correlations may or may  not be physically meaningful.

In as much as Milky Way GMCs have been the foundation of GMC studies, observations of these sources are affected by a number of challenging phenomena, mainly the blending of emission from multiple clouds along the line of sight. External galaxies offer an opportunity to study GMCs and star formation in different environments, including different metallicities and different galaxy types, and to make comparisons with our own Galaxy.  With the emergence of modern (sub)millimeter interferometers and large single-dish telescopes, it has become possible to resolve individual GMCs in nearby galaxies \citep{Schruba2017}.

GMCs are traced by emission from the low rotational (J) states of the CO molecules, which are excited via collisions at temperatures ranging from $\mathrm{5-20~K}$ \citep{Van1988}. A number of high resolution CO observations have been done in external galaxies, including M33 \citep{Engargiola2003,Rosolowsky2003,Rosolowsky2007,Gratier2012} and NGC 300 \citep{Faesi2018}. More recently, the Physics at High Angular resolution in Nearby GalaxieS project has mapped CO(2-1) emission from multiple galaxies, resolving the molecular gas reservoir into individual GMCs across the full disc \citep{Schinnerer2019}.

M33 is a flocculent spiral galaxy in the Local Group. It is metal poor but gas rich and has a metallicity of $\mathrm{12+log(\frac{O}{H})=8.36\pm 0.04}$ \citep{Rosolowsky2008A}. It is at a distance of $\mathrm{840~kpc}$ \citep{Freedman1991,Kam2015} and an inclination of $56\degr$ \citep{Kam2015}, which allows us to resolve gas components with minimum contamination along the line of sight and to map their inner structure of GMCs. Earlier studies of GMCs in this galaxy include those by \citet{Wilson1997,Rosolowsky2007,Tosaki2007,Miura2010,Gratier2010,Gratier2012,Tabatabaei2014}.

 The giant {H\,{\normalsize \text{II}} region} (GHR) NGC 604 is located in the northern arm of M33. This region has attracted interest because it has the highest star formation rate in the entire galaxy \citep{Miura2012}. The GHR has been observed in radio emission \citep{Viallefond1992,Wilson1992,Churchwell1999,Tosaki2007,Miura2010}, optical emission \citep{Drissen1993} and X-ray emission \citep{Tullmann2008}. Based on these previous studies, the H$\alpha$ nebula has a core-halo structure extending out to $\mathrm{200-400~pc}$. It contains more than 200 O-type stars that are surrounded by photoionized filaments and shells \citep{Relano2009}.
 
 \citet{Faesi2018} notes that determining true physical signatures in extragalactic studies is made difficult due to the wide range of source finding techniques and deferring observational characteristics (angular, spectral, and sensitivity) used. Hence, new analysis techniques are needed to overcome these problems and to understand the universality of Larson's relations.
 
 In this work, we present a dendrogram analysis of ALMA observations of $^{13}$CO(J=1-0) and 104 GHz continuum emission from NGC~604. These observations have better resolutions and sensitivities compared to prior observations, which helps to overcome many of the issues highlighted by \citet{Faesi2018}.  We look at whether the $^{13}$CO emission from NGC 604 obey Larson's relations in the same way as the $^{12}$CO emission from the same region as presented by \citet[][hereafter WS92]{Wilson1992}.  Using these new data, we measure the properties of the clouds and examine the state of the star formation in the region, and we compare to results presented earlier by \citet{Miura2010}. We present the observations and data reduction process in Section \ref{sec:Data}, the structure decomposition analysis and measurements of the cloud properties in Section \ref{sec:analysis}, and the results in Section \ref{sec:results}.  We discuss our results in Section \ref{sec:Discussion} and summarize our results in Section \ref{sec:Conclusion}.


\section{Observations and Data Reduction}
\label{sec:Data}

We use archival ALMA Band 3 observations of the $^{13}$CO(J=1-0) (110.27 GHz) line emission from NGC~604 obtained during Cycle 2 (project code 2013.1.00639.S; PI: T. Tosaki). The target was observed with the ALMA 12m array on 18 January 2015 for a total of 60 minutes on-source.  ALMA was in configuration C34-2/1 with 34 antennas (although two are flagged as unusable) arranged with baselines ranging from 15~m to 349~m, which yields a minimum beam angular resolution of 2.2~arcsec and a maximum recoverable scale of 29~arcsec (at 110.27 GHz). This corresponds to physical scales of 9 to 116 pc at the distance of $\mathrm{840~kpc}$ to M33. The observed field of view is 43~arcsec. J2258-2758 was used as a bandpass calibrator, Mars as a flux calibrator and J0237+2848 as a phase calibrator.

Four spectral windows were used in the observations.  Three of the spectral windows cover the $^{13}$CO (J=1-0) at 110.2~GHz, C$^{18}$CO(J=1-0) at 109.8~GHz  and  CH3OH at 96.7~GHz lines; each of these spectral windows contained 180 channels with widths of 244.14~kHz, covering a bandwidth of 117.2~MHz.  The fourth spectral window covered continuum emission from 98.56 - 99.50~GHz using 3840 channels with widths of 244.14~kHz ($\mathrm{\sim~0.664~kms^{-1}}$). Only the $^{13}$CO (J=1-0) and continuum emission are detected in this data.

The Common Astronomy Software Application package \citep[CASA; ][]{McMullin2007} version 5.6.1 was used to process the data.  We first performed the standard pipeline calibration on the visibility data and then produced line cubes and continuum images using  {\sc tclean}.  We set the pixel scale for both the continuum and line images to 0.36~arcsec.  The channel width for the $^{13}$CO image was set to 0.664 km~s$^{-1}$.  We used Briggs weighting with the robust parameter set to 0.5 to improve the angular resolution of the final images without severely compromising the image sensitivity.  The synthesized beam sizes are 3.2~$\times$~2.4~arcsec for the line data and 3.9~$\times$~2.8~arcsec for the continuum data.  The achieved rms sensitivity in the line data is 2.6~mJy~beam$^{-1}$ and continuum is 0.04~mJy~beam$^{-1}$.  The calibration uncertainty is expected to be 5\% \citep{Braatz2020}.

\begin{figure*}
	\includegraphics[width=\textwidth]{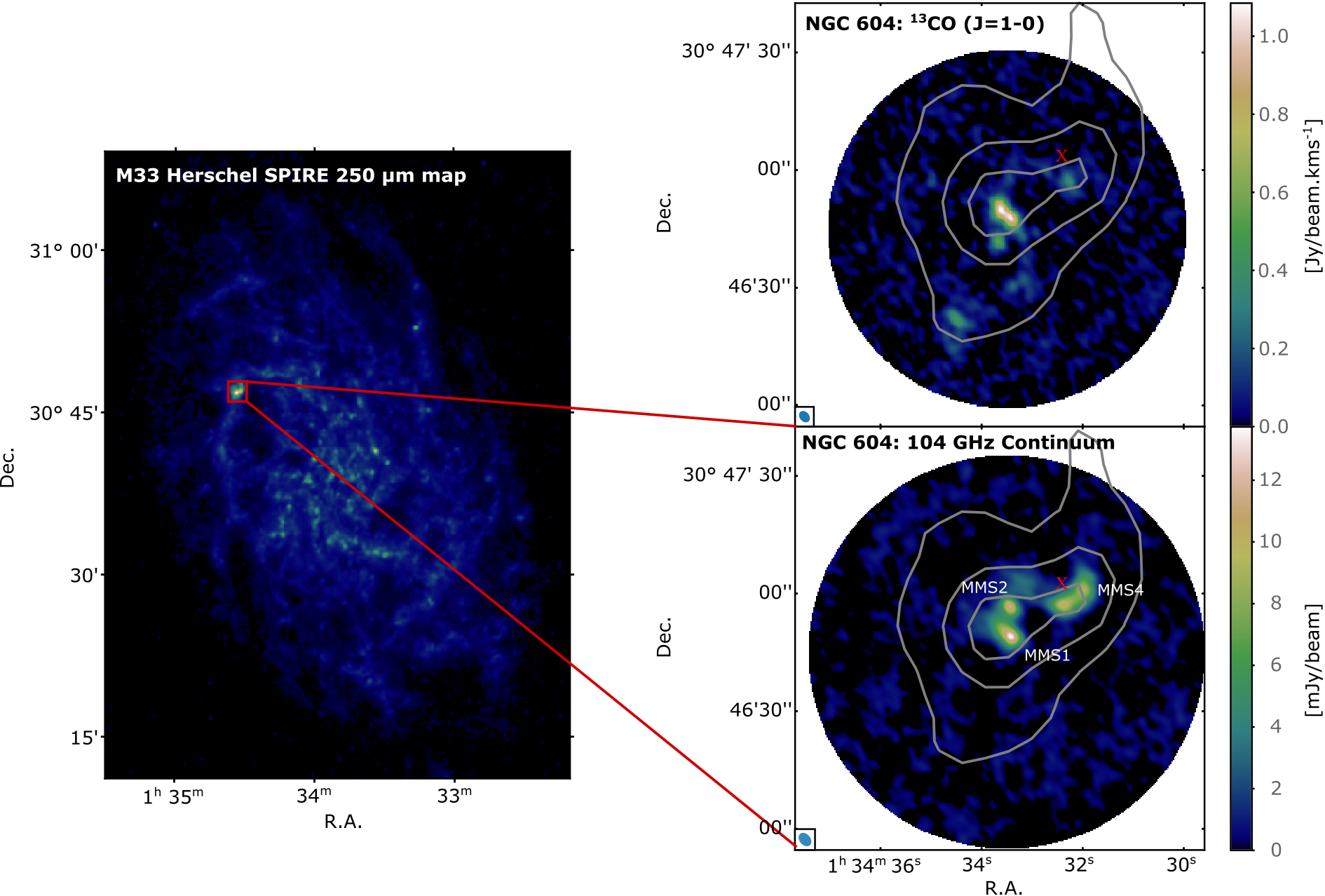}
    \caption{\textit{Left panel:} A 250~$\mu$m image of M33 tracing cold interstellar dust emission. \textit{Right top panel:} The $^{13}$CO(J=1-0) emission in NGC 604 as observed by ALMA.  \textit{Right bottom panel:} The ALMA 104~GHz continuum emission in NGC~604 resolved into three sources, which we call millimeter sources (MMS). The gray contours in both right panels show the $250~\micron$ emission, and the red cross symbol shows the centre of the GHR.} 
    \label{fig:Integrated_emission}
\end{figure*}

\begin{figure}
\begin{center}
    \includegraphics[width=\columnwidth]{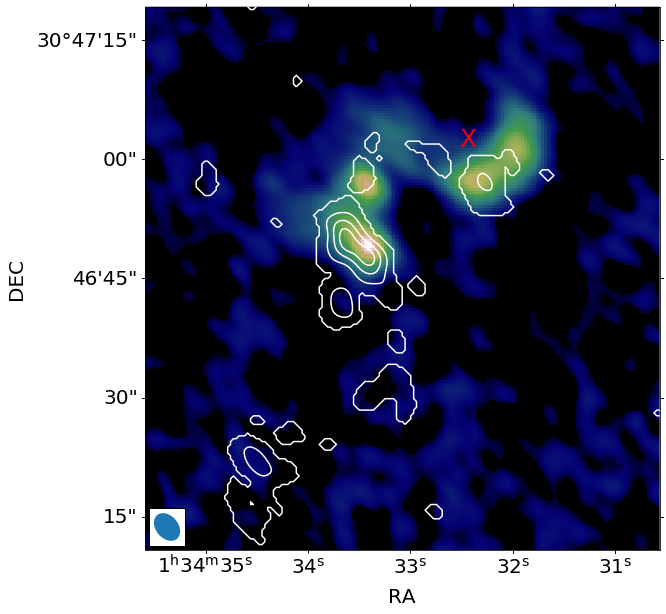}
\caption{{\bf The } ALMA 104~GHz continuum image of NGC 604 in colour with the integrated $^{13}$CO(J=1-0) emission overlaid as white contours. The contour levels represent 20, 40, 60, and 80\% of the peak emission. The angular resolution is $3''.9 \times 2''.8$ for ALMA 104~GHz continuum. The continuum emission is seen only near the centre of the GHR, and some regions with $^{13}$CO(J=1-0) emission do not have continuum emission. The color bar is the same as the bottom right panel of Fig \ref{fig:Integrated_emission}.}
\label{fig:ALMA continuum}
\end{center}
\end{figure}

The $^{13}$CO(J=1-0) integrated intensity map and the 104~GHz continuum map are shown in Figure \ref{fig:Integrated_emission}. As an additional visualization aid, the $^{13}$CO(J=1-0) emission is overlaid as contours on the continuum image in Figure \ref{fig:ALMA continuum}.

The 104~GHz continuum emission detected in NGC~604 (as shown in the bottom right panel of Figure \ref{fig:Integrated_emission}) is believed to be dominated by free-free emission (as indicated by the spectral energy distribution analyses of other galaxies by \citealt{Peel2011}, \citealt{Bendo2015}, and \citealt{Bendo2016}) that originates from OB stars within NGC~604.  We find spatial offsets between $^{13}$CO line and 104~GHz continuum emission as shown in Figure \ref{fig:ALMA continuum}. See Section \ref{sec:Discussion} for more details on their distribution.

\section{Structure Decomposition and Cloud Properties}
\label{sec:analysis}
\begin{figure}
\begin{center}
    \includegraphics[width=\columnwidth]{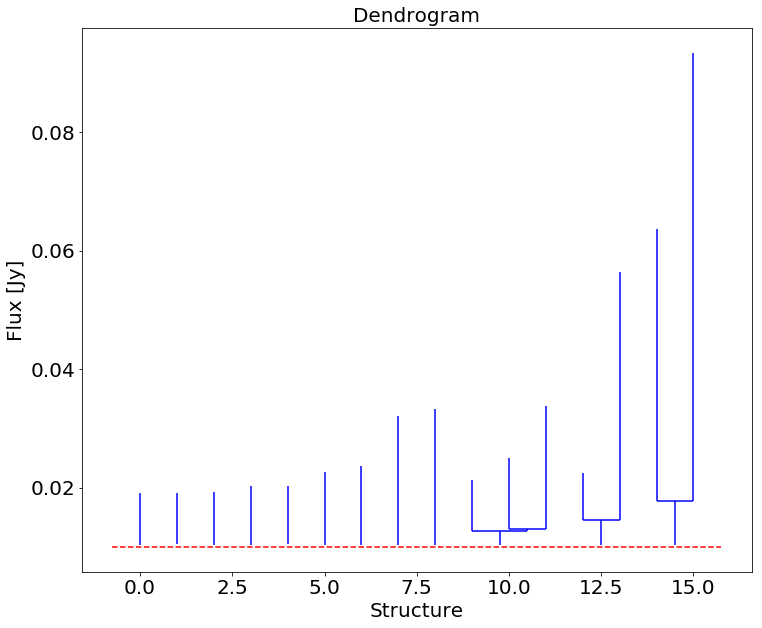}
\caption{The dendrogram of the ALMA $^{13}$CO(J=1-0) structures in NGC~604. The top of each vertical line indicates a leaf node, which we assume to be a molecular cloud. The horizontal red dotted line represents the minimum value of the tree, which is at 4$\mathrm{\sigma}$ noise level.}
\label{fig:dendrogram}
\end{center}
\end{figure}

\begin{figure*}
\begin{center}
    \includegraphics[width=\textwidth]{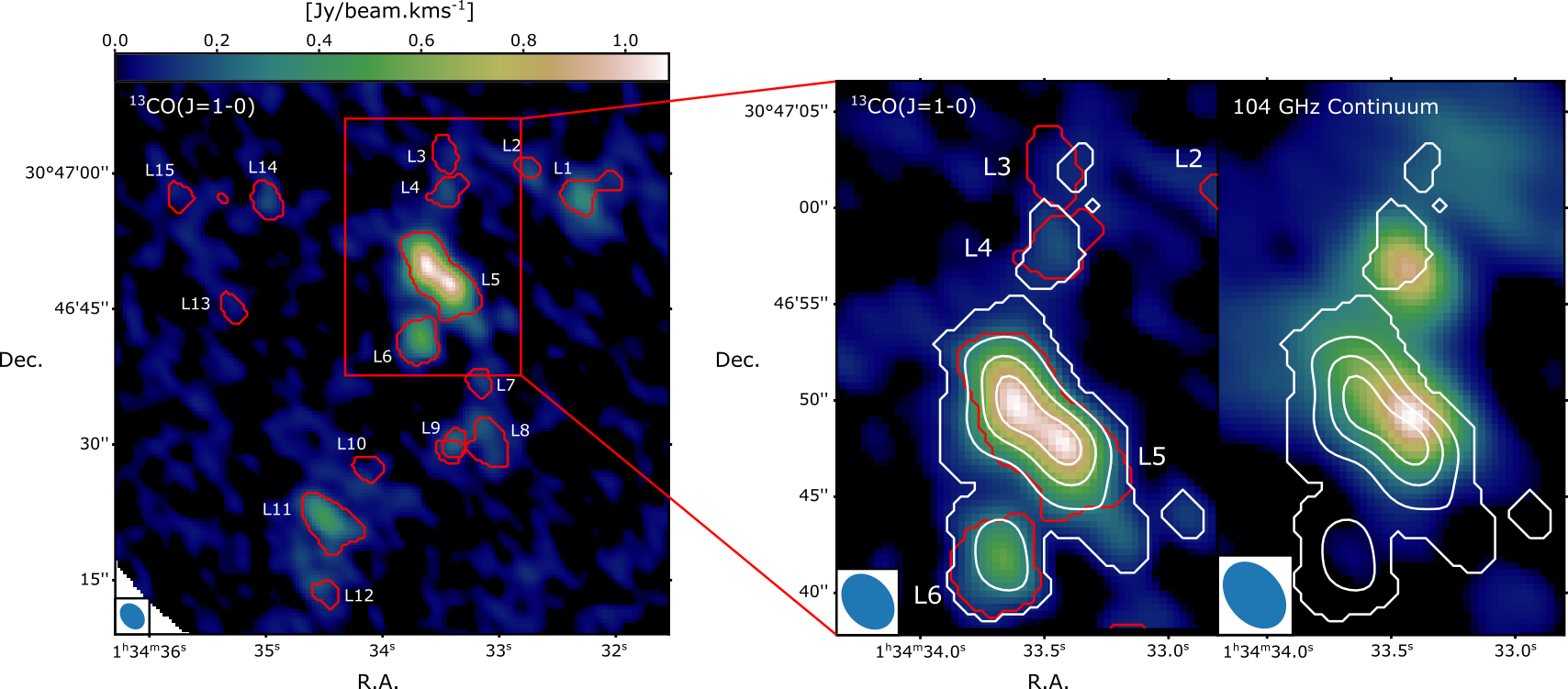}
\caption{The left panel shows the $^{13}$CO(J=1-0) emission from NGC604 with the red contours demarcating the clouds identified by astrodendro. The red box shows the NMA-8 region, which is shown in detail in the two right-hand zoomed panels. The right-hand zoomed panel shows the $^{13}$CO(J=1-0) emission from the four resolved molecular clouds (details as the larger amp), while the left-hand zoomed panel shows the 104~GHz continuum emission. White contours showing the $^{13}$CO(J=1-0) line emission are overlaid on both zoomed panels. The contour levels represent 20, 40, 60, and 80\% of the peak emission.}
\label{NMA8}
\end{center}
\end{figure*}

To identify structures within the $^{13}$CO(J=1-0) image cube, we used the \textsc{astrodendro} package, which decomposes emission into a hierarchy of nested structures \citep{Rosolowsky2008,Colombo2015}. This dendrogram technique, provide a precise representation of the topology of star forming complexes. Parameters were chosen so that the algorithm could identify local maxima in the cube above the 4$\sigma_{rms}$ level that were also 3$\sigma_{rms}$ above the merge level with adjacent structures. Isorsurfaces surrounding the local maxima were categorized as trunks, branches or leaves based on whether they were the largest contiguous structures (trunks), intermediate in scale (branches) or had no resolved substructure (leaves). The resulting dendrogram for $^{13}$CO(J=1-0) in NGC~604 is shown in Figure \ref{fig:dendrogram}. We identified 20 structures in the entire dendrogram, consisting of 15 leaves and 4 branches, using the above parameters. Spectra for the peak brightness pixels, for each leaf, are presented in Appendix \ref{Spectra}. We use letter {\bf L} to represent the leaf number in our labels for the structures. From now onwards, we shall refer to these leaves as molecular clouds.

We compared our results with the results from \citet{Miura2010}, who show observations of $^{12}$CO(J=1-0) line emission from NGC 604 as observed by the Nobeyama Millimeter Array. We detected and resolved the clouds that they labelled NMA 4, 7, 8, 9, 10. We, however, are not able to detect NMA 1, 3, 6, 11 and 12 above our 4$\sigma_{rms}$ noise level. This is because \citet{Miura2012} used a lower detection threshold of 3$\sigma$.  If we lower our detection threshold to 3$\sigma$, we can detect these sources, but we also detect additional spurious noise in the maps.  Given this situation, we chose to use only sources detected at the higher threshold. NMA 2 and 5 are outside of our field of view. We proceed to determine the basic properties of the identified structures at this point.

The basic properties of the identified structures are also determined by \textsc{astrodendro} using the bijection approach \citep{Rosolowsky2008}.  We extracted the molecular cloud properties using the approach described by \citet{Wong2017}. These properties include spatial and velocity centroids ($\mathrm{\Bar{x},\Bar{y},\Bar{v}}$), the integrated flux F, the rms line-width $\mathrm{\Delta v}$ (defined as the intensity-weighted second moment of the structure along the velocity axis), the position angle of the major axis $\phi$, and the scaling terms along the major and minor axes, $\sigma_{maj}$ and $\sigma_{min}$. From these basic quantities, we calculated additional cloud properties; these are listed in Table \ref{tab:Cloud Properties}.  The rms spatial size $\sigma_r$ is given by the geometric mean of $\sigma_{maj}$ and $\sigma_{min}$.  The spherical radius $R$ is set to 1.91 $\sigma_r$ following \citet{Solomon1987} and \citet{Rosolowsky2006}.  The luminosity-based mass for  $^{13}$CO(J=1-0) is computed using
\begin{equation}\label{eq:1}
   \mathrm{ \dfrac{M_{lum}}{M_{\odot}}  = \dfrac{X_{{}^{13}CO}}{2\times 10^{20}[cm^{-2}/(K~km~s^{-1})]}\times 4.4 \dfrac{L_{^{13}CO}}{K~km~s^{-1}~pc^2}\\
    = 4.4X_2 L_{^{13}CO}}
\end{equation}
from \citet{Rosolowsky2008}, where $\mathrm{X_{^{13}CO}}$ is the assumed $\mathrm{^{13}CO(1-0)-to-H_2}$ conversion factor. This calculation includes a factor of 1.36 to account for the mass of helium. Changes to the first term or conversion factor are represented with the parameter $\mathrm{X_2}$. We have adopted $X_2 = 5$ based on the average $\mathrm{^{13}CO(1-0)-to-H_2}$ conversion factor of $\mathrm{1.0\times 10^{21}cm^{-2}/(K~km~s^{-1})}$ for nearby disc spiral galaxies found by \citet{Cormier2018}. This average is equivalent to what would be expected for the conversion factor for a galaxy with $\mathrm{12+log(O/H)=8.4}$. This is close to the abundance of $\mathrm{12+log(O/H)=8.45\pm 0.04}$ measured for NGC 604 \citep{Esteban2009}. The scatter in $\mathrm{X_{^{13}CO}}$ value is 0.3 dex \citep{Cormier2018}. This uncertainty means that masses will have a systematic error of about a factor of 2.

The virial mass of molecular clouds derived assuming virial equilibrium is
\begin{equation}
  \mathrm{M_{vir}=189 \Delta v^2 R \hspace{1cm} [M_{\odot}]} \label{eq:2}
\end{equation}
where $\mathrm{\Delta v}$ is the linewidth in $\mathrm{km~s^{-1}}$ and R is the spherical radius in pc. This formulation assumes a truncated power-law density distribution of $\mathrm{\rho \varpropto R^{-\beta}}$ with $\beta = 1$ and with the assumption that magnetic fields and external pressure are negligible \citep{Solomon1987}. In this equation, $\mathrm{M_{vir}}$ is only defined for finite clouds with resolved radii.

The average molecular gas surface density $\mathrm{\Sigma_{lum}}$ is defined as
\begin{equation}
    \mathrm{\Sigma_{lum}=\dfrac{M_{lum}}{\pi R^2}} \hspace{1cm}  [\mathrm{M_{\sun} /pc^2}] \label{eq:3}
\end{equation}
where $\mathrm{ M_{lum}}$ is the luminosity-based mass.

The dynamic state of a cloud is described by the the virial parameter $\alpha_{vir}$ which is given by
\begin{equation}
    \mathrm{\alpha_{vir}= \dfrac{189\Delta v^2 R}{M_{lum}}}\label{eq:4}
\end{equation}
Allowing for uncertainties in measured parameters, a virtual ratio of $\le2$ is generally taken to mean that a cloud is gravitational bound. However, a cloud with an $\alpha_{vir}$  ratio significantly lower than this would need additional internal support (e.g. magnetic fields) to survive for longer than the usual dynamical timescale \citep{Faesi2018}.

The uncertainties in the molecular clouds properties R, $\mathrm{\Delta v}$, $\mathrm{L_{^{13}CO}}$ and $\mathrm{M_{lum}}$ are computed using a bootstrap method with 50 iterations. The bootstrapping determines errors by generating several trial clouds from the original cloud data. The properties are measured for each trial cloud, and the uncertainties are estimated from the variance of properties derived from these resampled and remeasured datasets. The final uncertainty in each property is the standard deviation of the bootstrapped values scaled by the square root of the oversampling rate. The bootsrap method is described in detail by \citet{Rosolowsky2006} and \citet{Rosolowsky2008}. Other uncertainties in derived properties presented in this work are calculated using the standard propagation of errors.


\section{Results}
\label{sec:results}
  \begin{table*}
    \centering{
    \caption{Cloud properties derived from $^{13}$CO(J=1-0) in NGC~604 using dendrogram analysis. See Section \ref{sec:analysis} for the details on how the properties were derived.}
    \label{tab:Cloud Properties}
    \resizebox{\textwidth}{!}{\begin{tabular}{cccccccccccc}
    \hline
        MC ID  & RA & DEC & $\mathrm{V_{LSR}}$ & $\mathrm{\Delta v}$ & $\mathrm{L_{^{13}CO}}$ & $\mathrm{R}$ & $\mathrm{M_{mol}}$ & $\mathrm{M_{vir}}$ & $\mathrm{\alpha_{vir}}$& $\mathrm{\Sigma_{lum}}$\\
       
        & J2000    & J2000  & $\mathrm{(km~s^{-1})}$ & $\mathrm{(km~s^{-1})}$ & $\mathrm{K~km~s^{-1}~pc^{2}}$ & (pc) & $\mathrm{(10^3~M_{\sun})}$ & $\mathrm{10^3~M_{\sun})}$ & & $\mathrm{M_{\sun}~pc^{-2}}$ \\ 
         \hline 
       
        L1  & $01^h 34^m 32^s.28$ & +30:46:57.07 & -245.7 & $2.4\pm 0.3$ & $498\pm 60$ & $9.8\pm 0.9$ & $11.0\pm 1.0$ & $10.5\pm 2.8$ & $1.0\pm 0.25$ & $36\pm 7$\\
       L2 & $01^h 34^m 32^s.73$ & +30:46:59.84 & -249.1 & $0.3\pm 0.01$ & $20\pm 3$ & $4.2\pm 0.6$ & $0.4\pm 0.06$ & $0.1\pm 0.0$ & $0.22\pm 0.04$  & $6\pm 2$\\
       L3  & $01^h 34^m 33^s.39$ & +30:47:01.85 & -243.8 & $0.7\pm 0.1$ & $60\pm 7$ & $5.7\pm 0.5$ & $1.3\pm 0.2$ & $0.6\pm 0.2$ & $0.42\pm 0.12$ & $13\pm 2$  \\
       L4  & $ 01^h 34^m 33^s.46$ & +30:46:57.98 & -244.4 & $1.3\pm 0.1$ & $78\pm 13$ & $6.9\pm 0.5$ & $1.7\pm 0.2$ & $2.1\pm 0.4$ & $1.2\pm 0.24$ & $11\pm 2$\\
        L5 & $01^h 34^m 33^s.54$ & +30:46:48.88 & -243.1 & $2.9\pm 0.3$ & $3660\pm 520$ & $13.4\pm 1.2$ & $80.5\pm 11.1$ & $21.3\pm 4.8$ & $0.3\pm 0.06$ & $143\pm 26$\\
        L6  & $01^h 34^m 33^s.67$ & +30:46:41.92 & -241.1 & $1.9\pm 0.2$ & $672\pm 97$ & $8.1\pm 0.6$ & $14.8\pm 2.0$ & $5.7\pm 1.3$ & $0.4\pm 0.1$ & $72\pm 11$\\
        L7  & $01^h 34^m 33^s.13$ & +30:46:37.09 & -252.0 & $1.4\pm 0.1$ & $122\pm 17$ & $8.5\pm 0.7$ & $2.7\pm 0.3$ & $3.0\pm 0.5$ & $1.1\pm 0.2$ & $12\pm 2$\\
        L8  & $01^h 34^m 33^s.16$ & +30:46:31.80 & -247.1 & $1.7\pm 0.2$ & $412\pm 51$ & $13.5\pm 1.2$ & $9.1\pm 0.9$ & $7.1\pm 1.8$ & $0.8\pm 0.2$ & $16\pm 3$\\
        L9  & $01^h 34^m 33^s.37$ & +30:46:30.44 & -252.4 & $0.8\pm 0.1$ & $47\pm 5$ & $5.3\pm 0.5$ & $1.0\pm 0.1$ & $0.7\pm 0.2$ & $0.7\pm 0.17$ & $12\pm 2$\\
        L10  & $01^h 34^m 34^s.18$ & +30:46:25.48 & -219.2 & $0.3\pm 0.03$ & $21\pm 3$ & $5.1\pm 0.5$ & $0.5\pm 0.06$ & $0.1\pm 0.02$ & $0.2\pm 0.04$ & $6\pm 1.1$\\
        L11  & $01^h 34^m 34^s.49$ & +30:46:21.91 & -220.5 & $2.2\pm 0.3$ & $1076\pm 158$ & $15.5\pm 1.8$ & $23.7\pm 4.0$ & $13.6\pm 3.6$ & $0.6\pm 0.17$ & $31\pm 7$\\
        L12  & $01^h 34^m 34^s.57$ & +30:46:14.66 & -217.9 & $0.5\pm 0.06$ & $32\pm 4$ & $5.0\pm 0.4$ & $0.7\pm 0.1$ & $0.2\pm 0.1$ & $0.3\pm 0.09$ & $9\pm 1.4$\\
        L13  & $01^h 34^m 35^s.30$ & +30:46:46.12 & -223.2 & $0.4\pm 0.07$ & $40\pm 6$ & $6.3\pm 0.6$ & $0.9\pm 0.1$ & $0.2\pm 0.1$ & $0.25\pm 0.08$ & $7\pm 1.4$\\
        L14  & $01^h 34^m 34^s.98$ & +30:46:57.35 & -229.8 & $0.8\pm 0.1$ & $114\pm 17$ & $6.6\pm 0.5$ & $2.5\pm 0.3$ & $0.8\pm 0.2$ & $0.31\pm 0.08$ & $18\pm 3$\\
        L15  & $01^h 34^m 35^s.80$ & +30:46:58.45 & -226.5 & $0.6\pm 0.08$ & $42\pm 6$ & $8.3\pm 0.7$ & $0.9\pm 0.1$ & $0.6\pm 0.2$ & $0.7\pm 0.19$ & $4\pm 1$\\
       
        \hline
   \end{tabular}}}
  \end{table*}
  
The properties of the the fifteen molecular clouds (leaves) identified by our dendrogram analysis are presented in Table \ref{tab:Cloud Properties}, and the left panel of Figure \ref{NMA8} shows the the locations of these clouds.  The two right panels in Figure \ref{NMA8} show magnified versions of the NMA-8 region.  \citet{Miura2010} only detected a single object in this region, but we detected four separate sources and resolved the structure in the brightest source. We discuss this more in Section \ref{sec:Discussion}.

\subsection{Scaling Relations}

\begin{figure}
\begin{center}
    \includegraphics[width=\columnwidth]{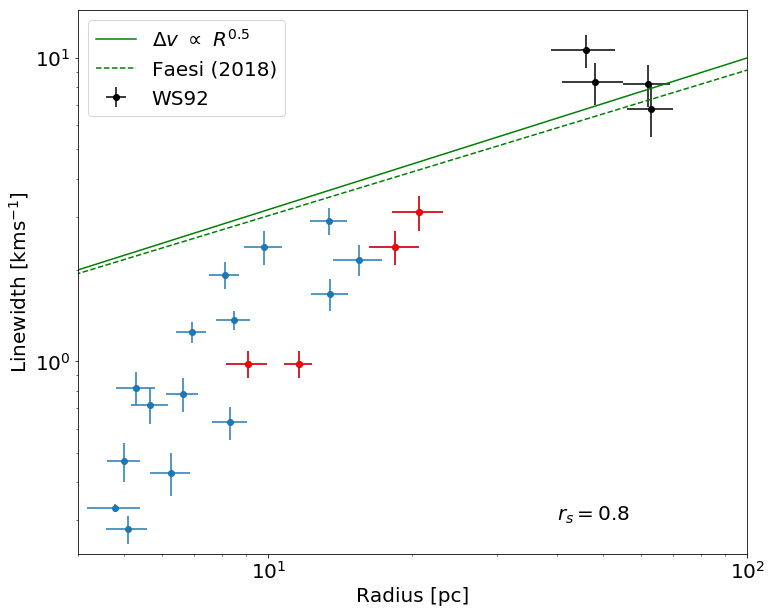}
    \caption{Size-linewidth relation of resolved molecular clouds in NGC~604. The green solid and dashed lines are the power-law slopes of Milky Way \citep{Solomon1987} and extragalactic \citep{Faesi2018} giant molecular clouds, respectively. The blue and red points represents the molecular clouds identified as leaves and branches in dendrogram tree, respectively. The black points are WS92 molecular clouds of NGC~604. There is a correlation with spearman rank of, $\mathrm{r_{s}=0.8}$.}
\label{size linewidth}
\end{center}
\end{figure}

Figure \ref{size linewidth} shows the size-linewidth relation for our sources. The clouds in blue are the fifteen clouds identified as unresolved substructure (leaves) by our analysis technique, and those in red are the branches which harbor resolved substructures. To investigate whether our molecular clouds are in virial equilibrium, we plotted molecular mass versus virial mass in Figure \ref{virial}.  In the absence of other forces, the virial parameter, which is the ratio of kinetic to gravitational potential energies, indicates the level of boundedness. The \textit{unbound} ones are those with $\mathrm{\alpha_{vir}> 2 }$,  while the \textit{bound}  are those with $\mathrm{\alpha_{vir} }$ between 1-2, and the ones with $\mathrm{\alpha_{vir} \leq 1}$ are in a state of forming stars.

\subsubsection{Size - Line width Relation}

The size-linewidth relation is commonly known as Larson's first law. The $\mathrm{\Delta v \varpropto R^{0.5}}$ relates the line-width in $\mathrm{km~s^{-1}}$ to the radius in parsecs \citep{Wong2017}. Large CO linewidths seen at parsec scales are evidence that these clouds are turbulent. It then follows from the size-linewidth relationship that there is a turbulent cascade of energy through the ISM \citep{Faesi2018} and that the form of this turbulence is described by its power law slope (1/2 for compressible, 1/3 for incompressible, \citep{McKee2007}). 

Figure \ref{size linewidth} shows the size-linewidth relation for our GMCs. We see that there is a clear trend, with larger clouds having larger linewidths, as is found in Milky Way clouds. The Spearman correlation coefficient for these data has the value of $\mathrm{r_{s}=0.8}$, which indicates that there is a correlation between size and linewidths of GMCs in NGC~604. We also show in Figure \ref{size linewidth} the Milky Way power-law slope (green solid line) from \citet{Solomon1987} and the extragalactic slope (green dashed line) from \citet{Faesi2018} for NGC~300. The relation for the NGC 604 clouds does not match the Milky Way and NGC~300 slopes; the linewidths at small radii for the NGC 604 data fall below the Milky Way and NGC~300 relations. In the figure, we plot results done by \citet{Wilson1992} (black points). Despite their results having considerable poor resolution, ($\mathrm{8~\arcsec~\times~7~\arcsec}$) compared to our ALMA $\mathrm{3.2~\arcsec~\times~2.4~\arcsec}$). There is consistency between the two results on large sizes having large linewidths (WS92 results) and smaller sizes having smaller linewidths (our clouds). The features are a typical characteristics of a turbulent spectrum which has a range of scales with increasing kinetic energy at large scales. We find their results to be in agreement with both the Milky Way and NGC~300 relations. \citet{Wong2017} and \citet{Wong2019} found a similar offset in the size-linewidth relationship between Milky Way and Large Magellanic Cloud data. They ascribed the discrepancy to two factors.  The first was the limitations in resolution of the Large Magellanic Cloud observations.  The second was the bijection approach in dendrogram analysis.  The rms linewidths in the dendrogram analysis tend to be underestimated for structures which are defined by high isocontour levels such as leaves because the full width of the spectral line is truncated by the isosurface boundary \citep{Rosolowsky2005,Rosolowsky2008}. NGC~604 and 30~Dor, one of the region \citet{Wong2017} studied, are both sites of massive star formation surrounding giant H\,{\normalsize \text{II}} regions and would both be places with high isocontour levels, so both of these locations could plausibly be affected by this truncation bias.  It is worth noting that other extragalactic studies have found no strong correlation between size and linewidth \citep{Colombo2014,Maeda2020}.

\subsubsection{Molecular Mass - Virial Mass Relations}

\begin{figure}
\begin{center}
    \includegraphics[width=\columnwidth]{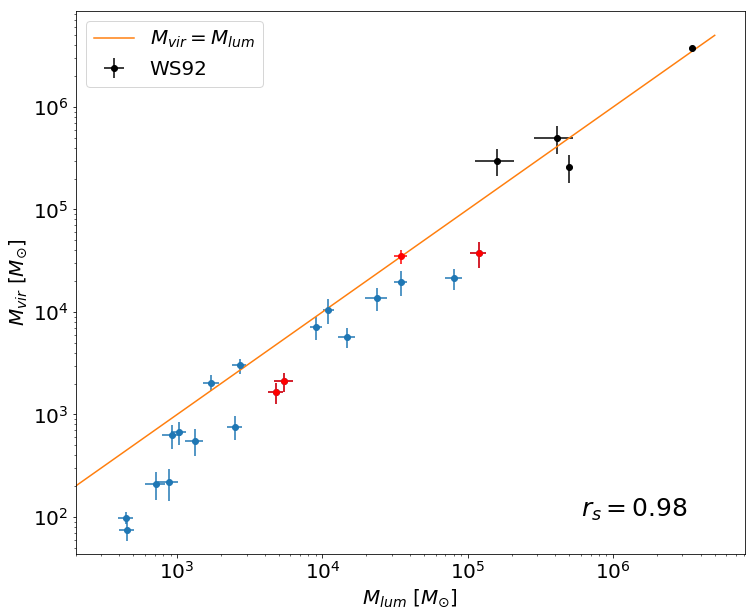}
    \caption{Luminosity mass plotted against virial mass. We see a strong correlation between these two parameters with a spearman coefficient of $\mathrm{r_{s}=0.98}$ indicated in the bottom right corner. The yellow line indicate a one-to-one relation. Despite being correlated most clouds fall below the  one-to-one relation. The red, blue and black points are the same as in Figure \ref{size linewidth}.}
\label{virial}
\end{center}
\end{figure}

The Milky Way observations have shown that the majority of GMCs are in self-gravitational equilibrium \citep[e.g.,][]{Larson1981,Solomon1987,Heyer2009,Heyer2015}. This leads to a direct correlation between $M_{vir}$ and the mass measured through other independent method (in our case the $^{13}$CO luminosity). Recent extragalactic studies of NGC~300 by \citet{Faesi2018} and NGC~1300 by \citet{Maeda2020} have found a strong correlation between $M_{vir}$ and $M_{lum}$ and a low scatter in $\alpha_{vir}$ near unity. We show in Figure \ref{virial} that the clouds in NGC~604 are in near virial equilibrium and that the data are strongly correlated, with a Spearman coefficient of $r_{s}=0.98$. Most of the clouds are lying below a one-to-one relation, illustrating that the masses estimated from the luminosities are slightly higher than the virial masses, which is a direct consequence of underestimating linewidths as discussed in the previous section. These clouds have virial parameters ranging from 0.2-1.1{\bf ,} indicating that some clouds are in virial equilibrium while others could be in a state of forming stars. The \citet{Wilson1992} data, which are also shown in Figure~\ref{virial}, largely seem consistent with the results from NGC 604.

\section{Discussion}
\label{sec:Discussion}

As seen in Figure \ref{fig:ALMA continuum}, both continuum and $^{13}$CO(J=1-0) emission are detected near the centre of the H\,{\normalsize \text{II}} region, although at locations further from the centre of the H\,{\normalsize \text{II}} region, we found a few locations with only $^{13}$CO(J=1-0) emission. The regions which are associated with continuum emission are actively forming stars.  We have labelled the three continuum sources with the abbreviation MMS (millimetre source), with MMS1 corresponding to L5, MMS2 corresponding to L4, and MMS4 corresponding to L1 as seen in the bottom right panel of Figure \ref{fig:Integrated_emission}.  \citet {Muraoka2020} also identified three sources in this region.  Our MMS2 corresponds to their MMS2, but they were able to resolve the brighter source, which we labelled as MMS1, into two sources labelled MMS1 and MMS3 (which is why we labelled our third source as MMS4). 

Regions only detected in $^{13}$CO(J=1-0) emission are dense molecular clouds with no active star formation. In these regions, atomic hydrogen (HI) could be forming H$_2$, and these clouds may form stars as the H\,{\normalsize \text{II}} expands. Previous studies in this region have found similar results and suggested that GMCs in NGC~604 are at different evolutionary stages, which would lead to sequential star formation induced by the expansion of GHR \citep{Tosaki2007,Miura2010}. To make comparison to the work done previously by \citet{Miura2010}, we use the nomenclature for their clouds and identify how many clouds we have resolved in each major GMC.

\subsection{NMA-8}

We have resolved NMA-8, the largest GMC in NGC~604 found by \citep{Miura2010}, into four individual molecular clouds that we labelled L3, L4, L5 and L6. It is possible that L5 contains two or more smaller clouds, but we could not separate them into smaller clouds when applying \textsc{astrodendro} to the $^{13}$CO data.  Based on the $^{12}$CO(J=1-0) observations, NMA-8 is known to be the most massive ($7.4\pm2.8 \times 10^5~M_{\sun}$) GMC in the GHR \citep[][and references therein]{Miura2010}.  Using $^{13}$CO(J=1-0), we estimate a virial mass of $\mathrm{0.8\pm 0.3\times 10^5~M_{\sun}}$ and a molecular mass of $\mathrm{1.2\pm 0.2\times 10^5~M_{\sun}}$ in NMA-8, which is a factor of 5 less than the  $^{12}$CO(J=1-0) molecular mass presented by \citet{Miura2010}. This is attributed to $^{13}$CO(J=1-0) only tracing the dense gas, hence, resolving away diffuse gas which make up large scale structure and also to the underestimation of linewidths. Our computed $^{13}$CO molecular mass for NMA-8 is comparable to the Orion A GMC, which has an estimated $^{12}$CO molecular mass of $\mathrm{1.1\times 10^5~M_{\sun}}$ \citep{Wilson2005}. The NMA-8 molecular mass estimate from $^{13}$CO is higher than the virial mass estimated from {\bf the} linewidths and the spherical radius but agree within the errors.  The estimated molecular mass of $\mathrm{0.8\pm 0.1\times 10^5~M_{\sun}}$ in L5 is comparable to Orion B in the Milky Way, which has a mass of $\mathrm{0.82 \times 10^5~M_{\sun}}$ \citep{Wilson2005}.

The association of L4 and L5 with 104~GHz continuum sources, which is expected to be dominated by free-free emission \citep[e.g. ][]{Peel2011, Bendo2015, Bendo2016}, clearly indicates that they are undergoing star formation. However, the peaks in the $^{13}$CO emission from these sources do no coincide exactly with the continuum peaks, as seen in the right zoomed panel of  Figure \ref{NMA8}. The continuum peaks lie closer to the centre of the {H\,{\normalsize \text{II}} region} than the $^{13}$CO peaks. This misalignment in this region has been reported previously by \citet{Miura2010}. The spatial offset between these peaks is an indication that these two tracers do trace different regions. The center has photoionizing stars which photoionize the gas surrounding which we trace by continuum in turn. The $^{13}$CO(1-0) line, being the lowest J-transition with a very low excitation temperature, preferentially traces cold dense molecular gas away from the centre. It is thus insensitive to the warm gas traced by the continuum emission. Earlier studies in NGC~604 by \citet{Muraoka2012} also found a temperature gradient in the NGC~604 clouds.

\subsection{Other GMCs in NGC 604}

We have for the first time resolved NMA-7 into three sources (L10, L11, and L12) and NMA-9 into three sources (L7, L8, and L9). Other than L1, these other GMCs are not associated with continuum sources and are not associated with ongoing star formation.   NMA-9 is the second massive and second largest complex in the imaged area, with a molecular mass of about $0.6\pm 0.1\times 10^5 M_{\sun}$. As we indicated before, the clouds without continuum emission could be places where the atomic gas is currently forming molecular gas, but when the GHR expands, these clouds may form stars. 

Generally, NGC~604 molecular clouds indicates that they are at different evolutionary stages within the {H\,{\normalsize \text{II}} region}, with some being associated with both continuum and line emission while others only line emission. Additional dendrogram analyses with higher resolution data will be necessary to explore these phenomena in more detail.


\section{Conclusions}
\label{sec:Conclusion}

We have presented ALMA $^{13}$CO(1-0) and 104~GHz continuum observations of NGC~604. Using the \textsc{astrodendro} algorithm, we identified 15 molecular clouds.  The main results are given as follows:

\begin{itemize}

    \item[1.] The identified molecular clouds have sizes $R$ ranging from 5-21 pc, linewidths $\Delta {v}$, of 0.3-3.0 $\mathrm{km~s^{-1}}$ and luminosity-derived masses $M_{lum}$, of $\mathrm{(0.4-80.5)\times 10^3~M_{\sun}}$. These sizes, linewidths and masses are comparable to typical Milky Way molecular clouds.
    
    \item[2.] For the first time, this work has resolved NMA-8, the most massive GMC, into four molecular clouds named L3, L4, L5 and L6, with L5 showing two clear peaks. We detect 104~GHz continuum emission from L5, although it is offset from the $^{13}$CO emission.
    
    \item[3.] We only detect 104~GHz continuum emission near the centre of GHR. Further out of the centre, only $^{13}$CO line emission is detected. This indicates that the GMCs in NGC~604 are in different evolutionary stages as previously suggested by \citet{Tosaki2007} and \citet{Miura2010}.   Additionally, we find a spatial misalignment between $^{13}$CO and 104~GHz continuum in NGC~604. The center has photoionizing stars which photoionize the gas surrounding which we trace by continuum in turn while the $^{13}$CO(1-0) line, being the lowest J-transition with a very low excitation temperature, preferentially traces cold dense molecular gas away from the centre. It is thus insensitive to the warm gas traced by the continuum emission. This is a confirmation of what previous studies found in the same region.
    
    \item[4.] We have found that the sizes and linewidths are correlated for the NGC~604 GMCs but that the relationship is offset from the Milky Way scaling relation. This may be a consequence of the limited resolution of our data or artefact of the dendrogram analysis as applied to bright sources. The relation for the clouds in NGC 604 is  consistent with the idea of compressible hierarchical turbulence in the ISM within this region. 
    
    \item[5.] We find a clear one-to-one relationship between virial mass and luminous mass indicating that the clouds in NGC~604 are in virial equilibrium. This relation is consistent with the earlier relation published by WS92.
    
    \item[6.] The virial parameter ranges from 0.2-1.1. This result entails that some of the molecular clouds are below $\mathrm{\alpha_{vir}}=1$ which means that not only are they in a state of forming stars but photoionizing stars have been formed. Other clouds have $\alpha_{vir}$ values near unity, which means that they are in virial equilibrium.

\end{itemize}
 

\section*{Acknowledgements}

We thank the anonymous referee for the helpful comments in improving this manuscript. Research reported in this publication was supported by a Newton Fund project, DARA (Development in Africa with Radio Astronomy), and awarded by the UK's Science and Technology Facilities Council (STFC) - grant reference ST/R001103/1. This work makes use of the following ALMA data: ADS/JAO.ALMA2013.1.00639.S. ALMA is a partnership of ESO (representing its member states), NSF (USA) and NINS (Japan), together with NRC (Canada), MOST and ASIAA (Taiwan), and KASI (Republic of Korea), in cooperation with the Republic of Chile. The Joint ALMA Observatory is operated by ESO, AUI/NRAO and NAOJ. This research made use of \textsc{astrodendro}, a Python package to compute dendrograms of Astronomical data \footnote{http://www.dendrograms.org/}. This research made use of \textsc{astropy},\footnote{http://www.astropy.org} a community-developed core Python package for Astronomy \citep{astropy2013, astropy2018}.
\\\\
\textbf{DATA AVAILABILITY}: The raw data underlying this article is available on the ALMA archive: ADS/JAO.ALMA2013.1.00639.S. The calibrated image data generated for this research will be shared on reasonable request to the corresponding author.




\bibliographystyle{mnras}
\bibliography{example} 





\appendix


\appendix \section{Peak Spectra for the sources}
Presented here in Figure~\ref{Spectra} are the spectra (as measured at the peak of the emission) for each of the molecular clouds identified in our dendrogram analysis.

\begin{figure*}
\begin{center}
    \includegraphics[width=\columnwidth]{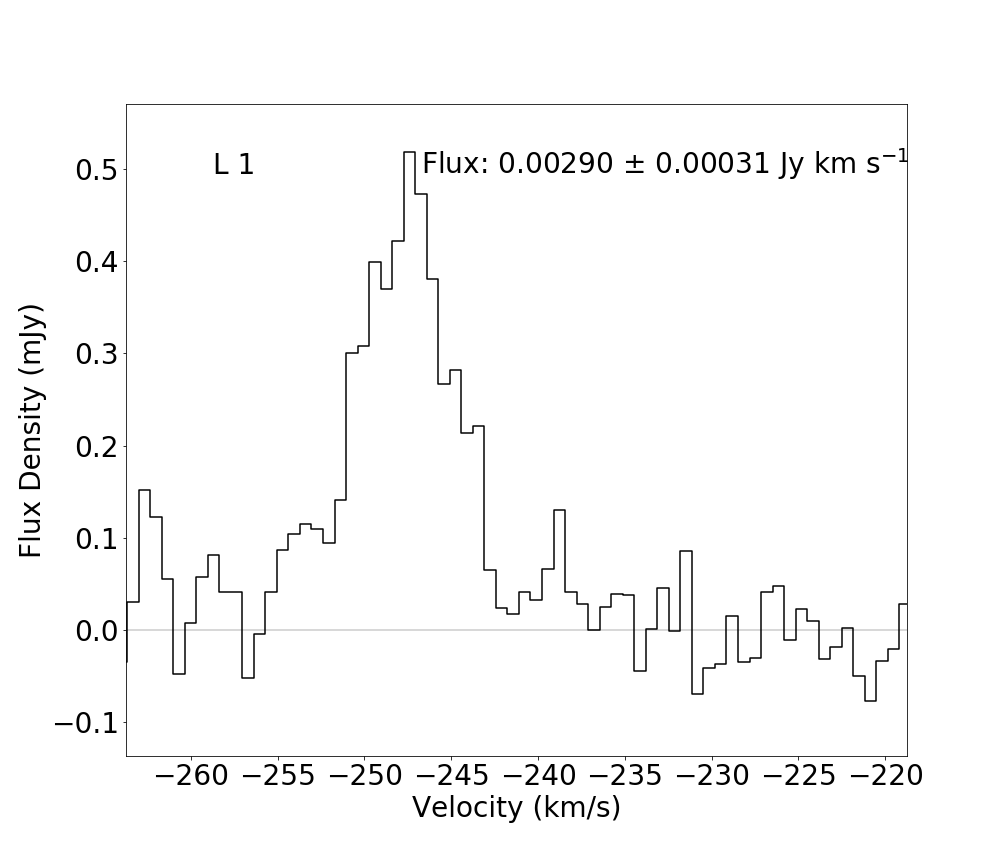}
    \includegraphics[width=\columnwidth]{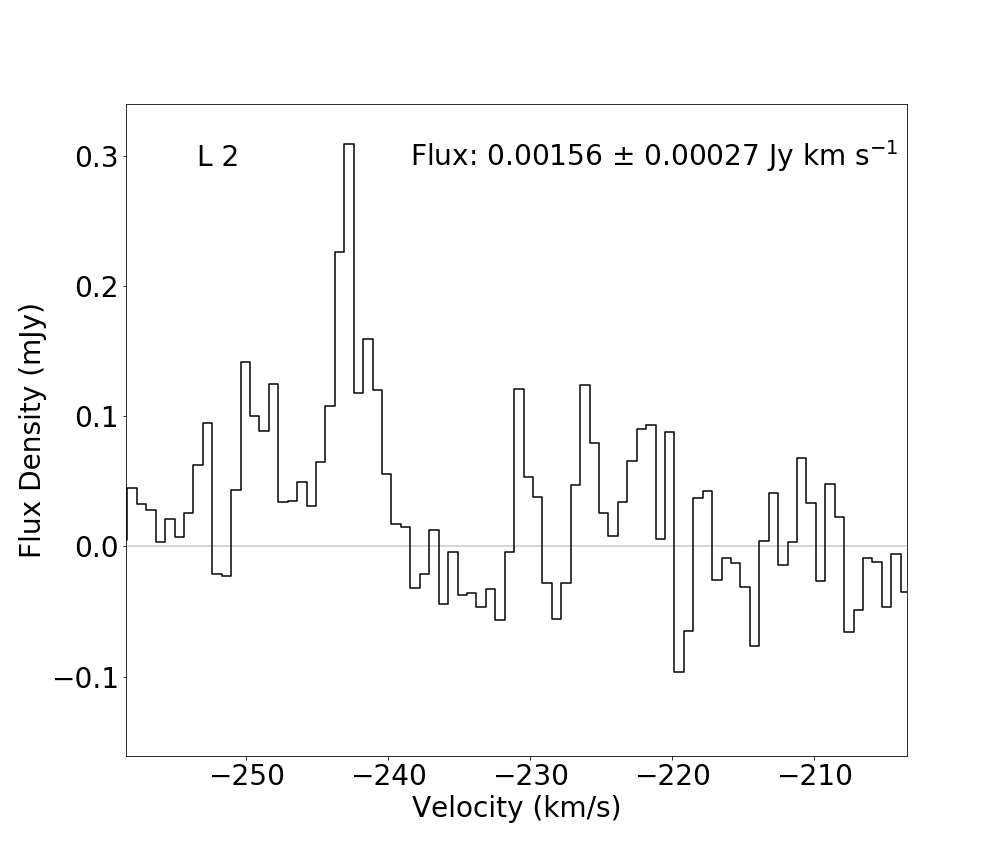}\\
    \includegraphics[width=\columnwidth]{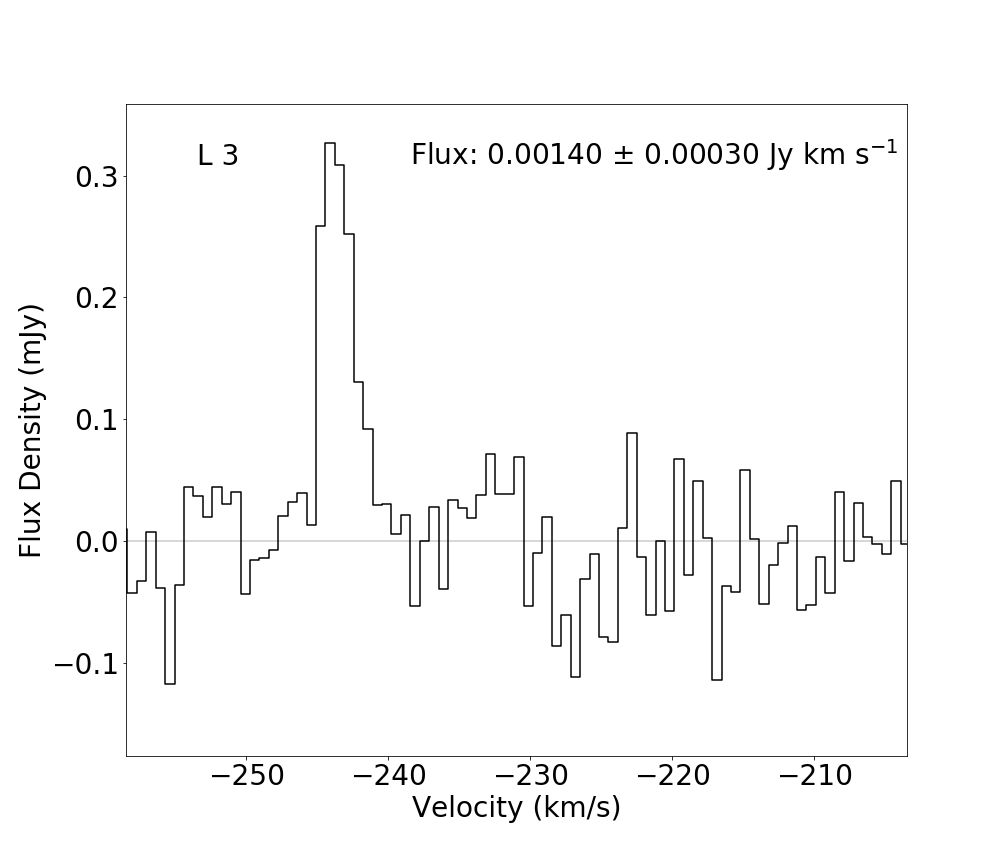}
    \includegraphics[width=\columnwidth]{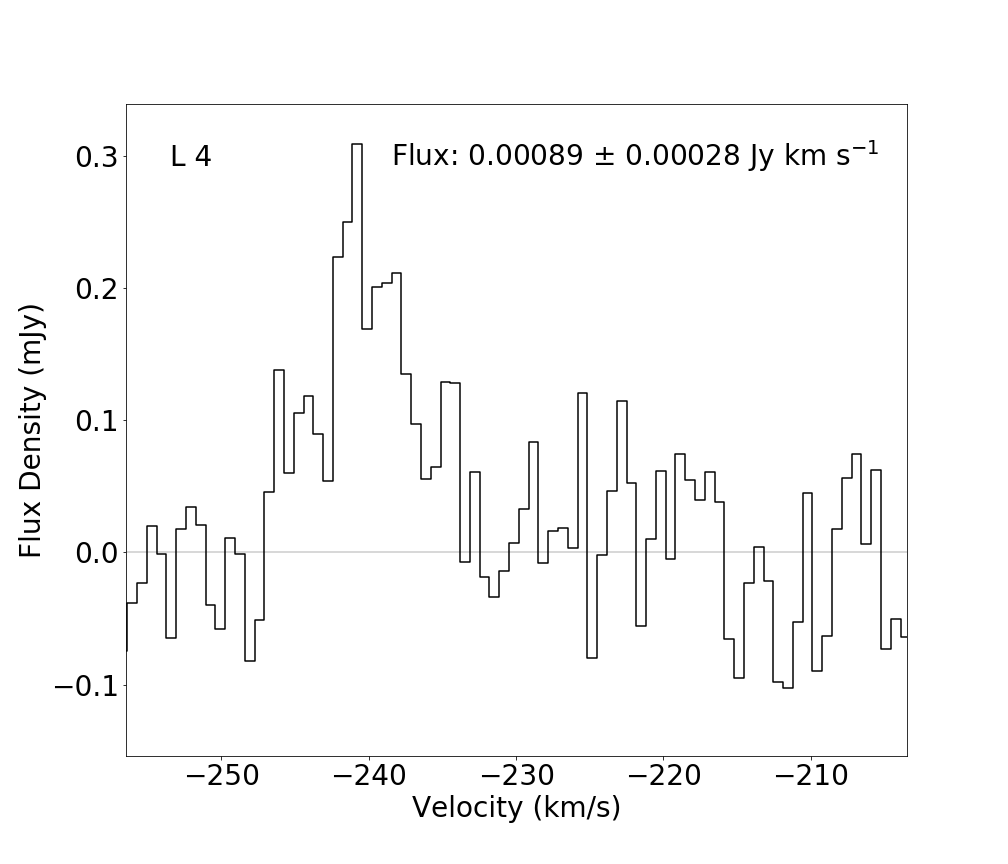}\\
    \includegraphics[width=\columnwidth]{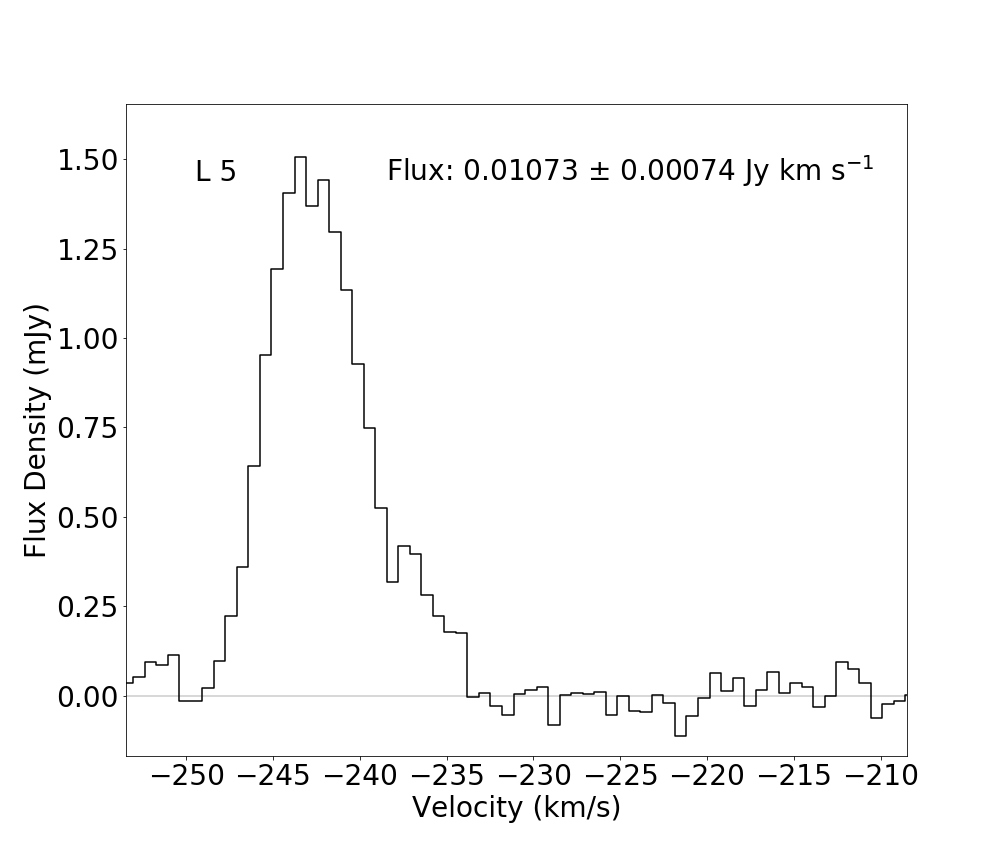}
    \includegraphics[width=\columnwidth]{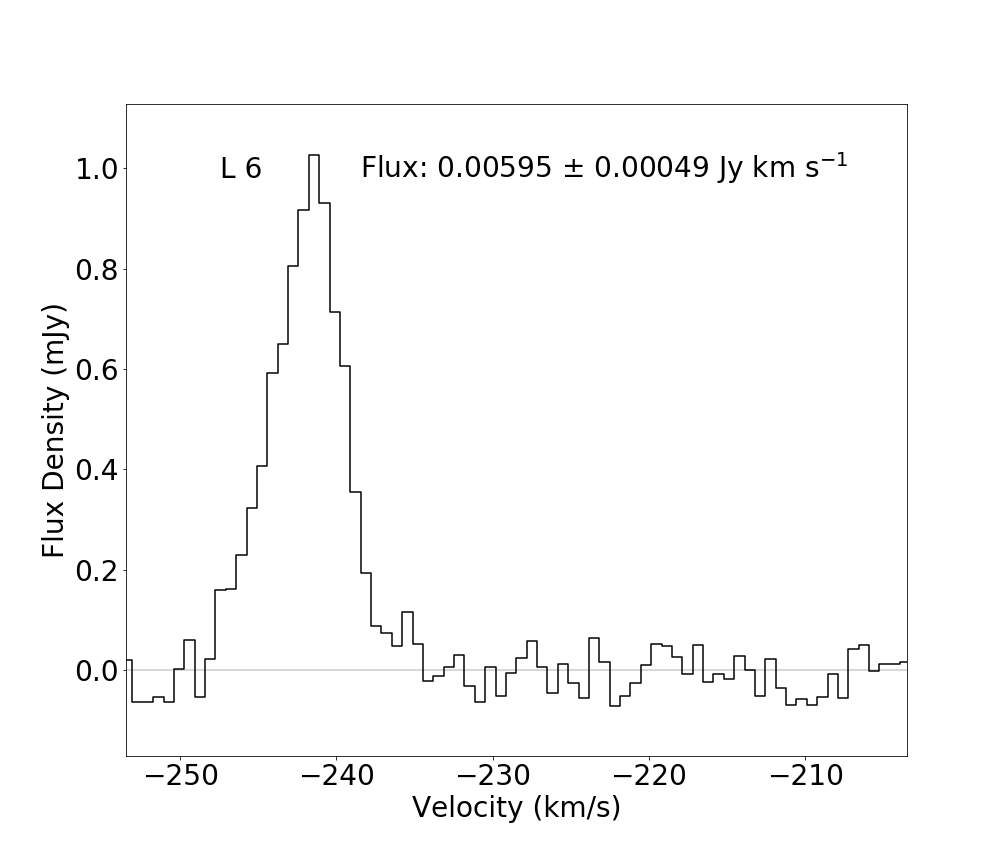}
\caption{NGC~604 GMC spectra as measured at the peak of the emission from each source.}
\label{Spectra}
\end{center}
\end{figure*}

\addtocounter{figure}{-1}
\begin{figure*}
\begin{center}
    \includegraphics[width=\columnwidth]{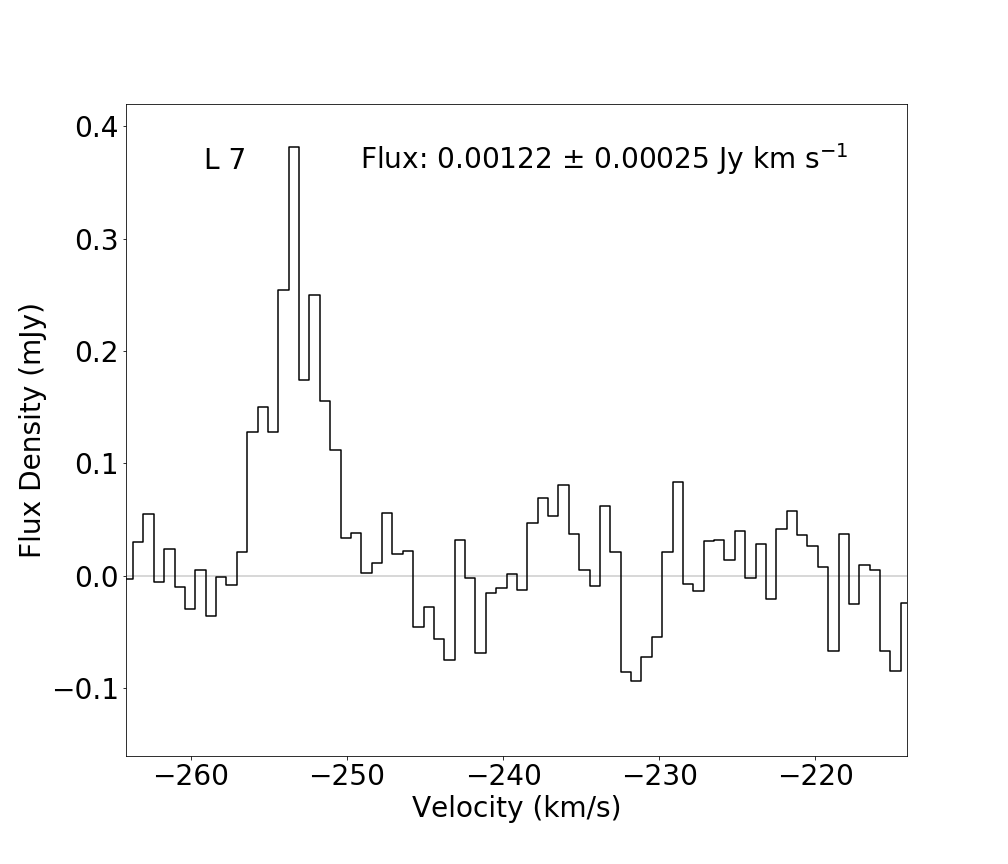}
    \includegraphics[width=\columnwidth]{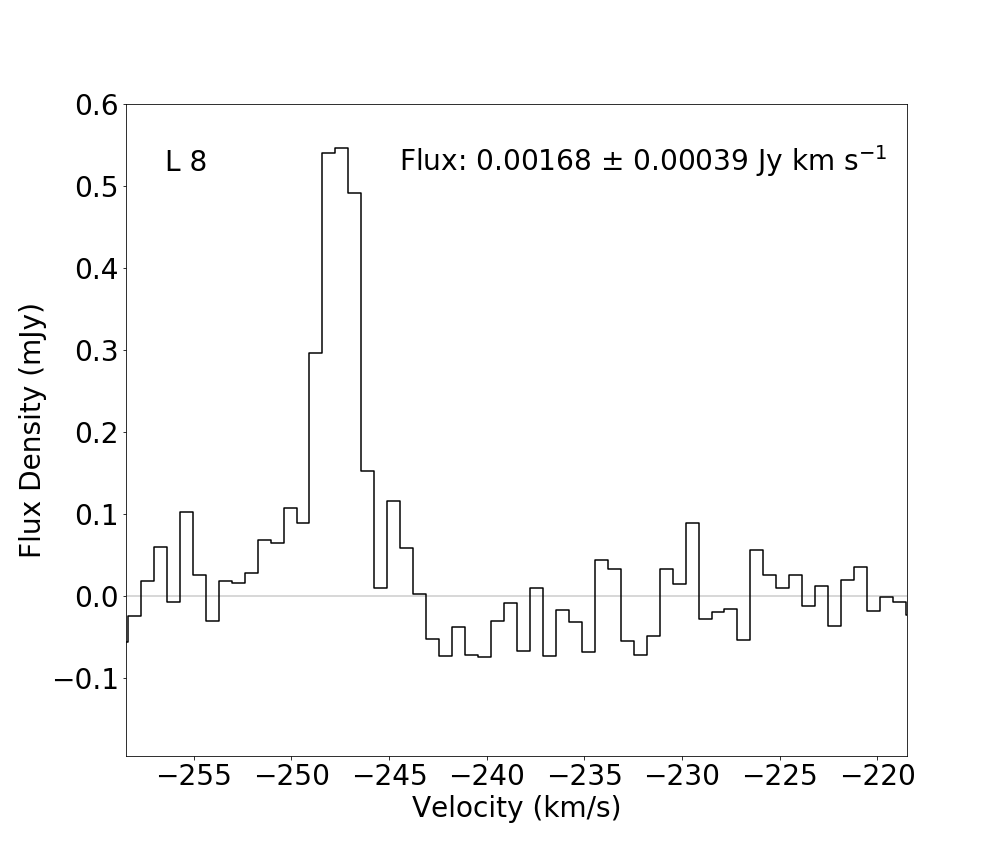}\\
    \includegraphics[width=\columnwidth]{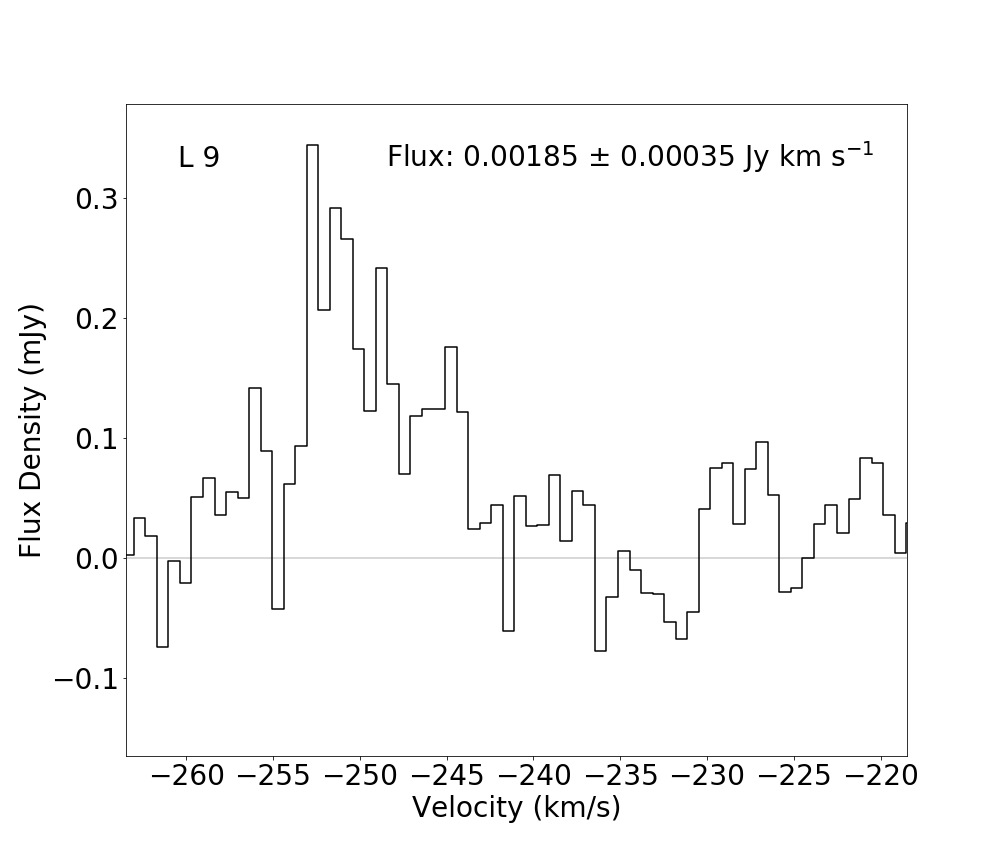}
    \includegraphics[width=\columnwidth]{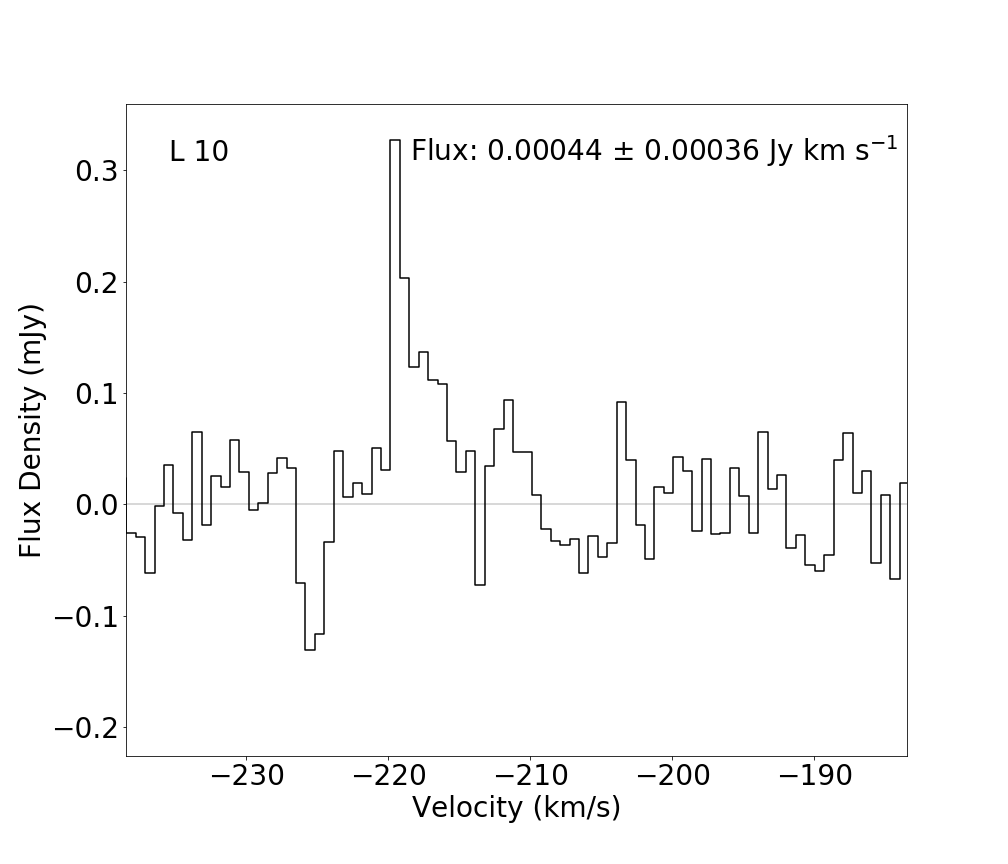}\\
    \includegraphics[width=\columnwidth]{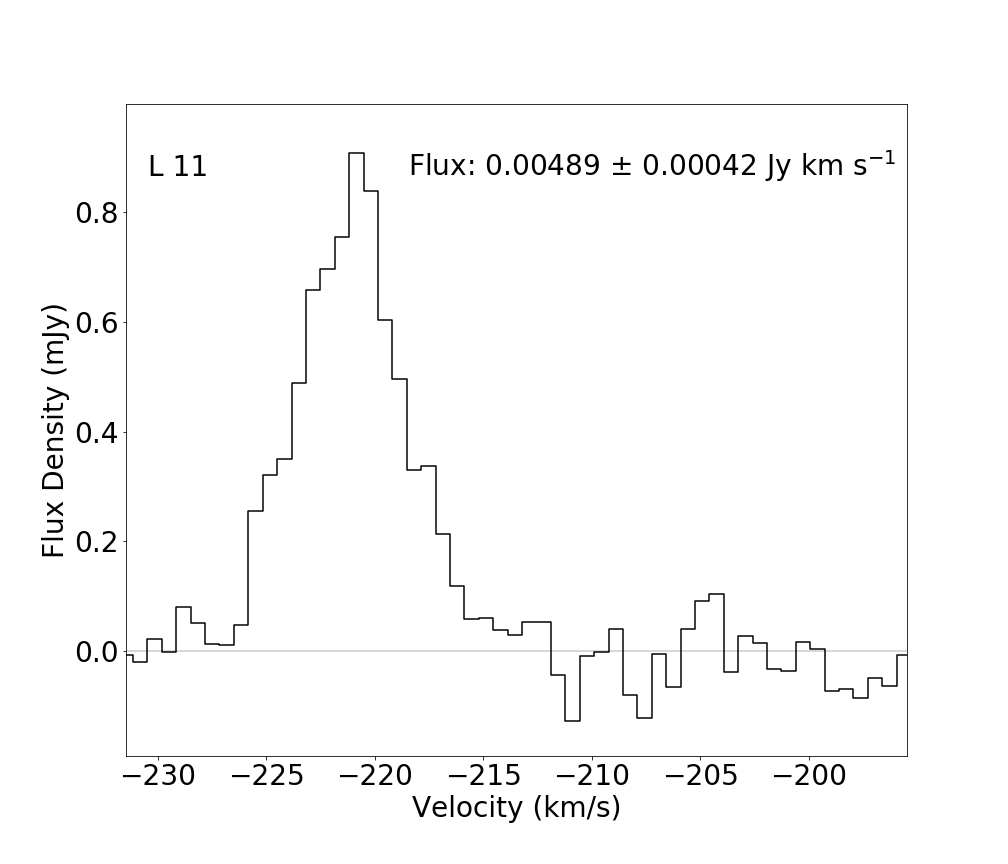}
    \includegraphics[width=\columnwidth]{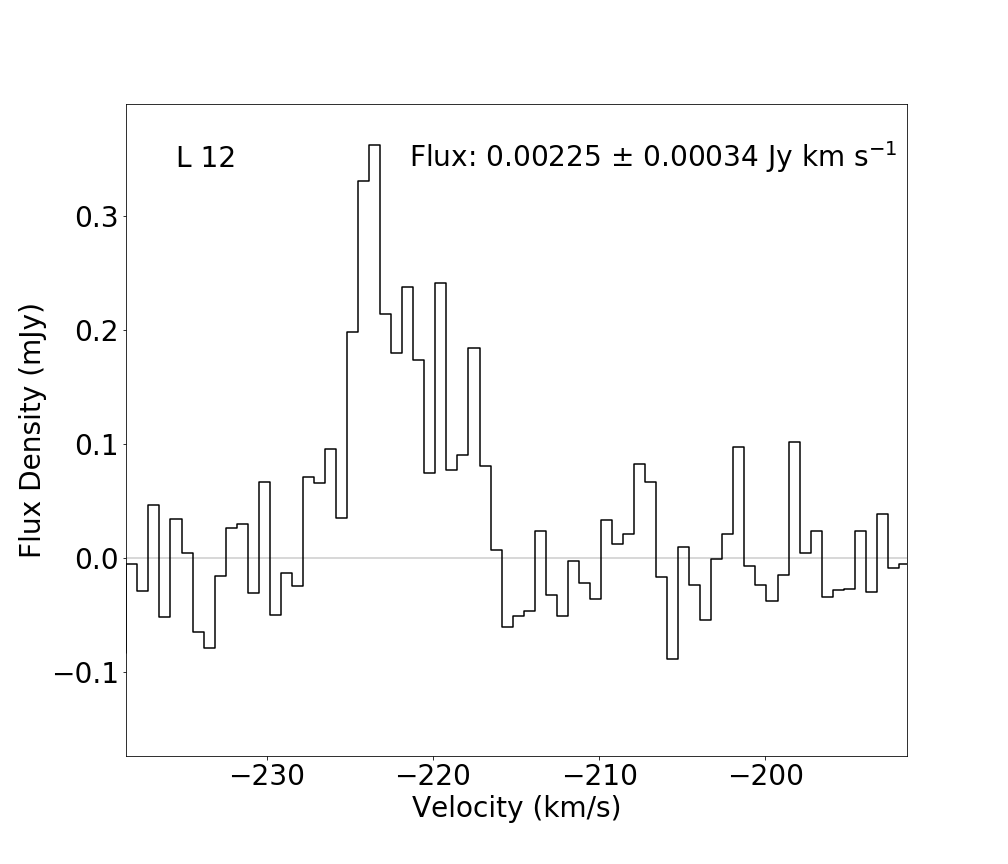}
\caption{continued.}
\end{center}
\end{figure*}

\addtocounter{figure}{-1}
\begin{figure*}
\begin{center}
    \includegraphics[width=\columnwidth]{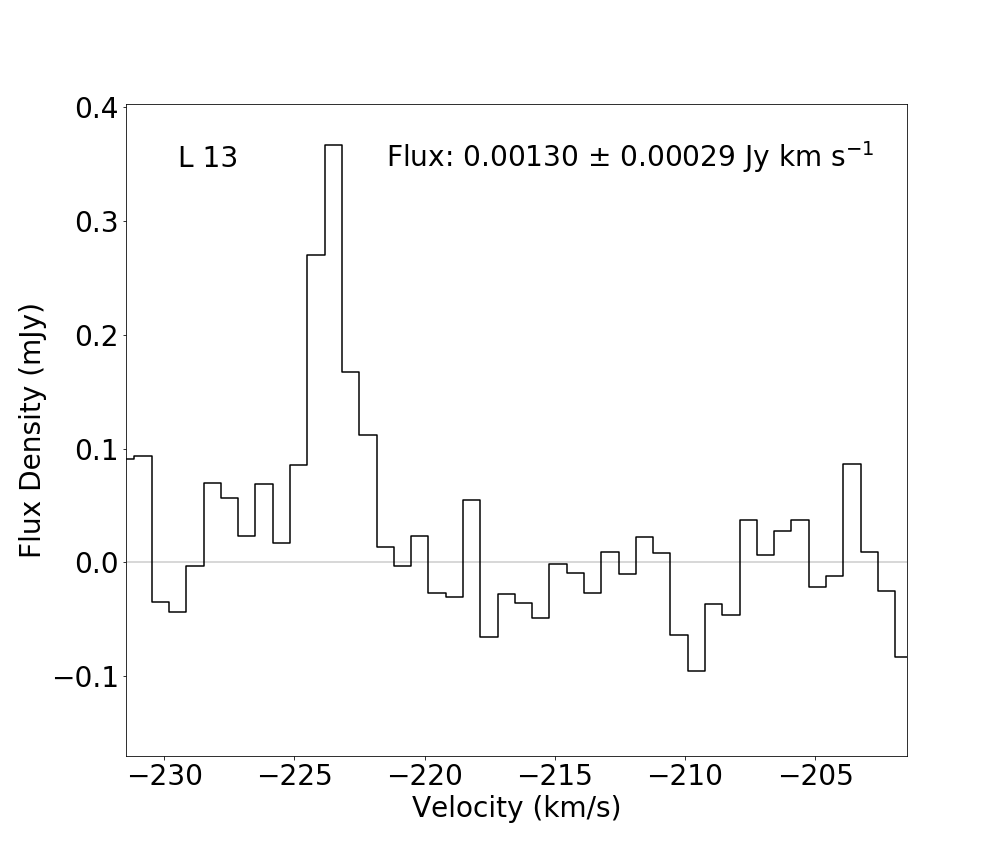}
    \includegraphics[width=\columnwidth]{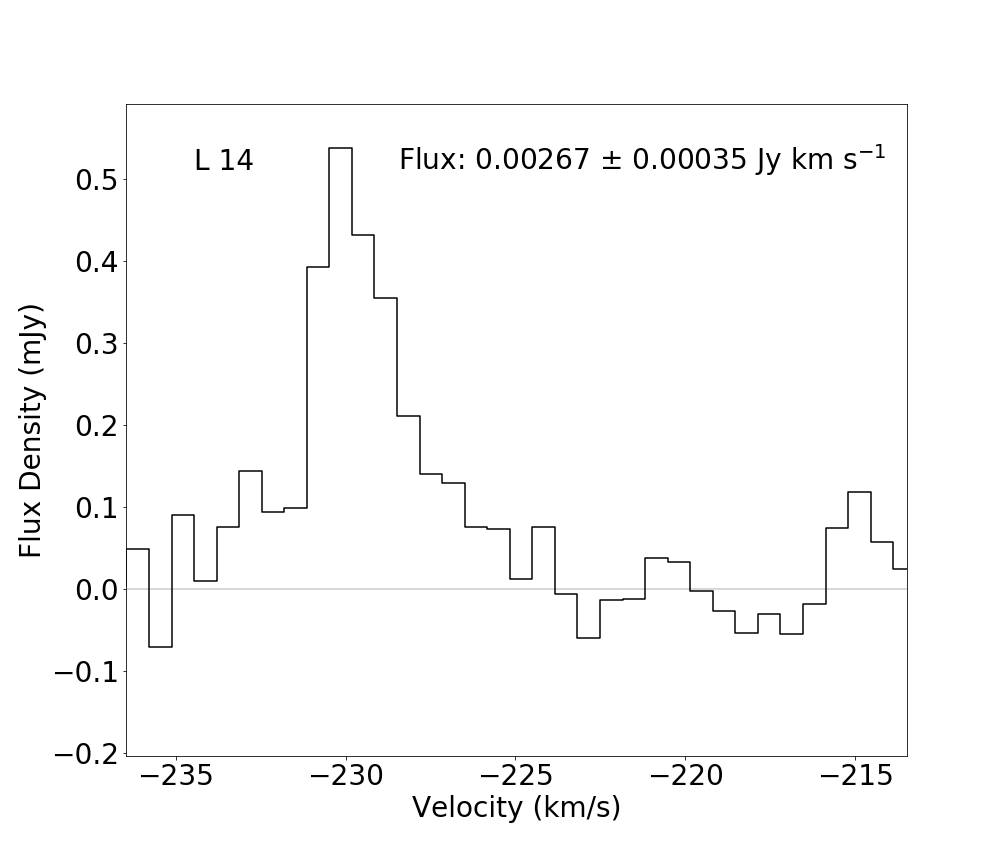}\\
    \includegraphics[width=\columnwidth]{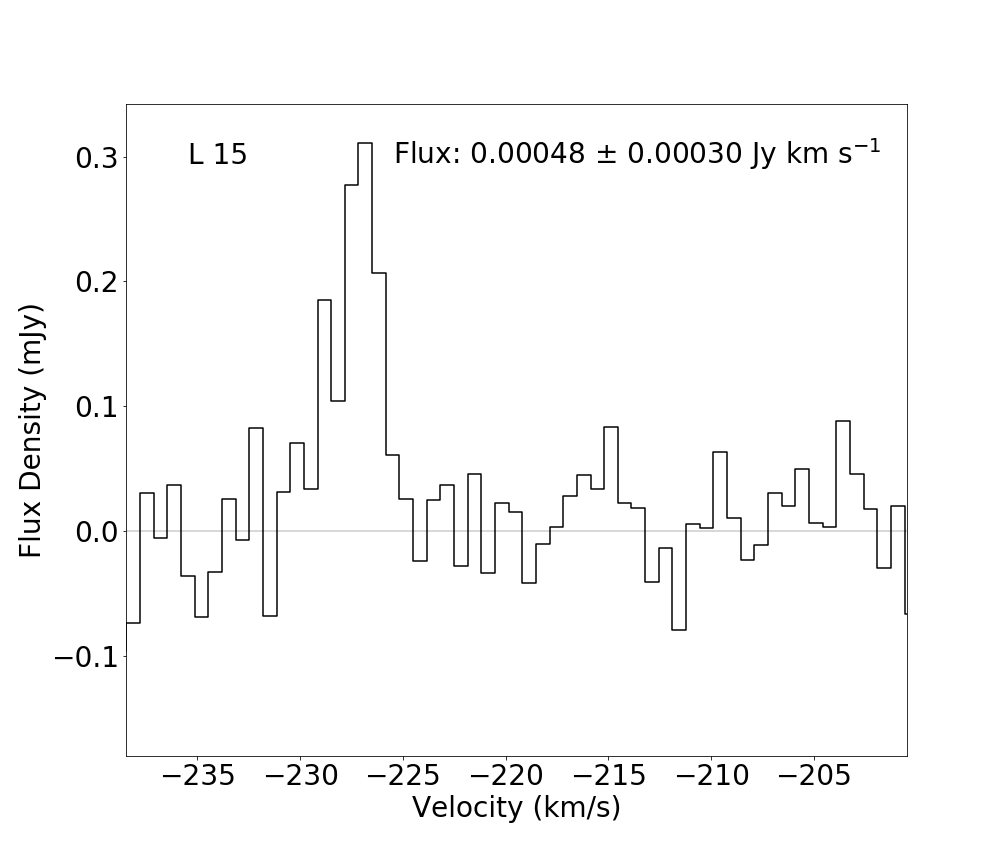}
\caption{continued.}
\end{center}
\end{figure*}


%
\label{lastpage}
\end{document}